# THE MESH-MATCHING ALGORITHM: AN AUTOMATIC 3D MESH GENERATOR FOR FINITE ELEMENT STRUCTURES

Béatrice COUTEAU *, Yohan PAYAN **, Stéphane LAVALLEE **

* INSERM U518, Pathologies ostéoarticulaires, C.H.U. Purpan, BP 3103, 31026 Toulouse cedex 3, France
** Laboratoire TIMC/IMAG, UMR CNRS 5525, Faculté de Médecine Domaine de la Merci, La Tronche, France

ADDRESS FOR CORRESPONDENCE
Yohan PAYAN
TIMC-IMAG
Pavillon Taillefer, Faculté de Médecine
38700 LA TRONCHE
Tel: (+33) 4 56 52 00 01
Fax: (+33) 4 56 52 00 55
e-mail: Yohan.Payan@imag.fr

ABSTRACT

    Several authors have employed Finite Element Analysis (FEA) for stress and strain analysis in orthopaedic biomechanics. Unfortunately, the use of three-dimensional models is time consuming and consequently the number of analysis to be performed is limited.

    The authors have investigated a new method allowing automatically 3D mesh generation for structures as complex as bone for example. This method called Mesh-Matching (M-M) algorithm generated automatically customized 3D meshes of bones from an already existing model. The M-M algorithm has been used to generate FE models of ten proximal human femora from an initial one which had been experimentally validated. The new meshes seemed to demonstrate satisfying results.

## 1. Introduction

Finite Element (FE) analysis of muscular-skeletal systems has been developed in order to assess strain and stress field within the different structures. This method presents a wide range of application domains as bone remodeling analysis (Huiskes et al., 1993, 1991; Weinans et al., 1991) , mechanical behavior of bones with or without an implant (Prendergast et al., 1990, Rubin et al., 1993; Skinner et al., 1994; Kuiper et al., 1996; Mann et al. 1995) and fracture process understanding (Vichnin et al., 1986; Lotz et al., 1991; Janson et al., 1993). These analysis require to know the exact geometry and mechanical properties of the different structures. Unfortunately, FE analysis are usually limited to only one specimen due to the prohibitive amount of manual labor required to generate a three-dimensional mesh. According to this reason, many two dimensional bone models have been developed in the orthopaedic research field, offering thus the possibility of high 2D mesh refinement, with a limited manual intervention (Brown et al., 1982; Carter et al., 1984; Rybicki et al., 1972; Weinans et al., 1988). In practice, when 3D finite element analysis for bone was carried out, some compromises were often made in terms of homogeneity (Valliappan et al., 1977; Oonishi et al., 1983), symmetry (Huiskes and Heck, 1981) or mesh refinement (Valliappan et al., 1977; Vichnin and Batterman, 1982). Moreover, 3D models, based on "average" bone geometry, have also been developed (Lotz et al., 1988; Weinstein et al., 1987), loosing thus any patient-oriented specificity. However, patient-specific three-dimensional FE models are important as they would be a way of correlating mechanical predictions with clinical results. This is the reason why the automated FE modeling of bone by using CT scan voxels as developed by Keyak et al. (1990) presents a great interest.

The objective of the present paper was to suggest a new method allowing automatical 3D mesh generation for structures with complex geometry as in case of bones. The Mesh-Matching (M-M) algorithm has been used to generate automatically customized 3D meshes of proximal femora from an existing 3D mesh.

## 2. Materials and Methods

The M-M algorithm is based on a registration method which was originally proposed for applications in computer-integrated surgery (Lavallée et al., 1995, 1996; Szeliski et al., 1996). This algorithm was applied here to infer an existing FE bone model to the same type of bone but from another patient. This paper focused more specifically on the proximal region of the femur (from the top of the head to the bottom of the lesser trochanter).

### 2.1 Mesh-Matching (M-M) algorithm.

This section presentes an outline of the elastic registration method which is used to deform an object in order to match another object (Szeliski and Lavallée, 1996).

The idea consists of finding a transform **T** which is the combination of a rigid-body transform $RT$, a global warping $W$ and a local displacement function $S$ built on a hierarchical and adaptive grid of displacements basis (*oct-tree splines*) :

$$\boldsymbol{T}_\mathbf{p} = RT \circ W \circ S \tag{1}$$

Where **p** is a vector gathering the 6 parameters that defines $RT$, the 12 to 30 parameters that defines $W$ and the thousands of local displacement vectors that defines $S$.

Let $\mathbf{M} = \{M_i, i = 1...N_1\}$ and $\mathbf{P} = \{P_i, i = 1...N_1\}$ be the sets of model and patient features, obtained by segmentation algorithms. The elastic registration algorithm minimizes a least-square criterion $E(\mathbf{p})$, given by :

$$E(\pmb{p}) = \sum_{i=1}^{N_1} \frac{1}{\sigma_i^2} \left[ dist\left(\pmb{P}, \pmb{T}_p(M_i)\right) \right]^2 + \pmb{R}_p \qquad (2)$$

Where **R** defines a regularization term which is applied to *S* in order to obtain a smooth displacement function.

$\sigma_i^2$ is the variance of the noise of the measurement *i* (Besl and McKay, 1992).

*dist* is the distance between the set **P** and a point M$_i$' (transformed by **T**). In that case, the distance *dist* was a 6-D distance function, as proposed by Feldmar and Ayache (1996).
The optimization of *E*(**p**) is performed by using the Levenberg-Marquardt algorithm (Press *et al.*, 1992) and a modified conjugate gradient algorithm in the hierarchical representation of **T**, in order to smooth the solution and to speed up the minimization

2-2 Finite Element meshing using the M-M algorithm

Transverse CT images (Siemens, DRH2) were performed on 11 cadaveric femora devoid of pathological signs. One millimeter thick slices were performed at 3 mm interval for the epiphyses and at 20 mm interval for the diaphysal region. Each image was subjected to an edge detection to extract bone contour lines. Each contour line was defined by a finite number of parametric cubic splines. The connection of cubic curves from two successive slices provided the 3D bone surface.

*Reference 3D mesh* - The so called "reference 3D mesh" was manually obtained from one of the femora by using hexaedric (8 nodes) and wedge (6 nodes) elements. The FE model comprised 3572 3D elements and it has been widely validated by means of experimental techniques. Mechanical properties of the bone were chosen from literature (Hobatho et al., 1991) in case of cortical bone and from predictive relationships with the Computed Tomography (CT) number (Couteau *et al.*, 1998) in case of cancellous bone. Firstly, a vibrational technique provided a good agreement between experimental resonant frequencies and those obtained numerically. Secondly, extensometric measurements demonstrated relatively low errors between experimental strains and those calculated numerically (Couteau, 1997).
The mesh generation is particularly difficult in the proximal region of the femur because element layer have to be, as far as possible perpendicular to the curved axis of the proximal femur (Vander Sloten and Van Der Perre, 1995). This is the reason why the automatic mesh generation was limited to the proximal femoral epiphysis. Consequently, the reference 3D mesh was defined by the elements located between the top of the head and the bottom of the lesser trochanter.

*Patient specific data sets* - The 3D surfaces of the other ten femora which were computed from CT scans were meshed with 2D elements (quads and triangles) in order to obtain the external points of the bone structure.
Then, the reference femur and each femur surface points were superimposed by using centroids of each data set as well as principal axis of inertia. The next step was the matching of the 3D reference mesh to the "target" points. (Fig.1) by means of the M-M algorithm.

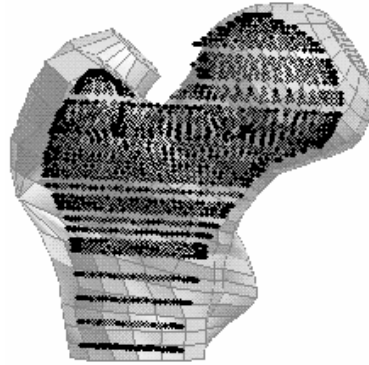

Fig. 1 : Superimposition of the reference 3D mesh (gray) with the 3D surface target points (black) from one of the ten others femora.

*M-M algorithm application* - The M-M algorithm computed first the volumetric function **T** (1) that transformed external nodes of the reference mesh into surface "target" points. This elastic volumetric registration took less than 30 seconds on a DEC Alpha 5000 workstation. Then this transformation **T** was applied to all the nodes of the 3D reference mesh leading to the new 3D mesh (this step required less than 10 seconds). Fig. 2 illustrates the mesh generation of one femur based on the reference mesh transformation.

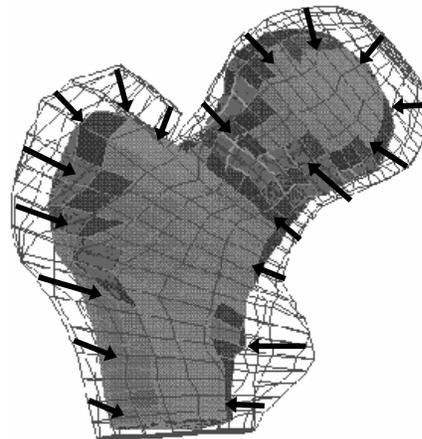

Fig. 2 : 3D mesh of the patient's femur (shaded gray) generated from the "reference" 3D mesh (wireframe).

## 3. Results

The application of the M-M method to the ten proximal femora demonstrated successful transformations. These have been tested by checking element distortion with respect to an ideal shape. The mechanical software (MSC/Patran V7.5) checked the distortion of each element through the angle between isoparametric lines of the element. Criterion of this test consisted in verifying whether the angle was greater than 45° or less than 135° in order to reduce the influence of the element distortion on the accuracy of the numerical integration (Dhatt et Touzot, 1984). If the angle was found outside this range of values a warning message was declared for the element. Our reference mesh had 14% of the elements with at least one angle outside the 45°-135° range. Nevertheless, the mean value of the worst angles of the declared distorted elements was equal to 35° and 75% of the distorted elements had their worst angle between 30° and 45°. This was not so far from the reasonable range and probably explained why the validation of the reference mesh was satisfactory. Concerning the ten others meshes which have been generated from the reference mesh, the rate of distorted elements was around 15% and the mean worst angle was equal to 34° (Fig. 3). The percentage of distorted elements with the worst angle between 30° and 45° was around 72%.

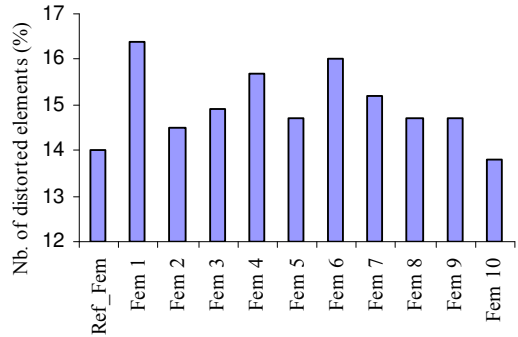

Fig. 3: Number of distorted elements in percentage of the total number of elements equal to 1758.

According to present results, the M-M method does not seem to make the mesh worse in spite of differences between femur geometries (Fig. 4.).

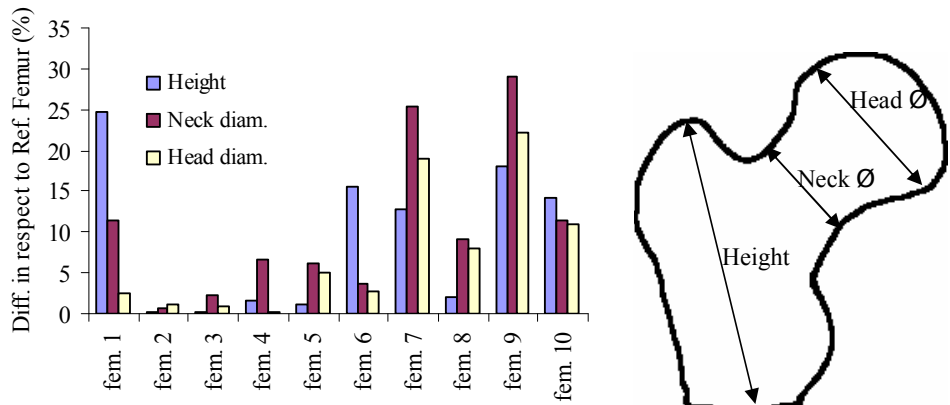

Fig. 4: Geometry differences in respect to the geometry of the reference femur. Dimensions of height, neck r and head diameter were given in % of the reference femur.

Actually, the differences between femur geometries have been quantified by comparing geometrical parameters (height, neck Ø and head Ø) of the ten femora with that of the reference femur. The maximum differences were around 25%, 30% and 23% for respectively the height, the neck and the head diameters.

## 4. Discussion
The M-M method allowed the automatic generation of FE meshes of patient proximal femora from an existing 3D model. For the ten patient femora tested in this paper, the element shape checking was satisfactory in comparison with the reference model. Therefore, the potential of the M-M method to mesh complex structures becomes evident. The necessity of having an initial mesh can be seen as a disadvantage of this method. Nevertheless, this can give the opportunity to have an optimal model with desired refinement mesh regions according to the geometric irregularities. Comparing with the automatic method introduced by Keyak et al. (1990), the advantage of the M-M method consists in the smooth surface representation which allows to get strains at specific points on the surface. Moreover, this method can be extended to all sorts of element (cubic, tetrahedral, ...). In the light of these results, limits of the method do probably exist but they were not yet reached there. Intrinsically, the elastic registration does not accept very different shapes to match. The next step will consist in assessing the limits of the M-M method by using it with very different sizes of basic shapes. Afterward, the M-M method will be tested with hollowed shape in order to take into account the whole parts of long bones.

To conclude, this paper has introduced a new method (Mesh-Matching algorithm) to automatically mesh 3D structures in the framework of mechanical analysis by the Finite Element Method. This method seems to show a great potential for the mechanical analysis of structures and it probably represents a new approach in the meshing or re-meshing techniques. Nevertheless, evaluation of the technique has to be performed on many kinds of bone structures in order to clarify the limits of the algorithm.
Finally, the originality of this work lies in the found link between two different disciplines belonging to the orthopaedic domain, i.e. computer-integrated surgery and mechanical analysis.

**Acknowledgments**
Richard Szeliski and Eric Bittar are acknowledged for their contributions on the elastic registration algorithm used in this paper. Marie-Christine Hobatho is acknowledged for her initial collaboration.